\documentclass[pre,twocolumn,a4paper,noshowpacs]{revtex4}
\usepackage{ifpdf}
\usepackage[usenames]{color}
\usepackage{graphicx}
\usepackage{amssymb}
\usepackage{amsmath}
\usepackage{subfigure}

\begin{document}

\title{Minimal model of associative learning for cross-situational lexicon acquisition}

\author{Paulo F. C. Tilles and Jos\'e F. Fontanari}
\affiliation{Instituto de F\'{\i}sica de S\~ao Carlos,
  Universidade de S\~ao Paulo,
  Caixa Postal 369, 13560-970 S\~ao Carlos, S\~ao Paulo, Brazil}

\begin{abstract}
An explanation for the acquisition of word-object mappings is the associative learning in a cross-situational scenario.
Here we present analytical results  of the performance of a simple associative learning algorithm for 
acquiring  a one-to-one mapping between $N$ objects and  $N$ words based solely 
on the co-occurrence between objects and words. In particular, a learning trial in our learning scenario consists of the presentation of $C + 1 < N$ objects together with a target word, 
which refers to one of the objects in the context. 
We find that the learning times are distributed exponentially  and the learning rates are  given by
$\ln{\left[\frac{N\left(N-1\right)}{C + \left(N-1\right)^{2}}\right]}$ in the case the $N$ target words are sampled randomly  and by $\frac{1}{N} \ln \left [ \frac{  N-1 }{C} \right ] $
in the case they follow a deterministic presentation sequence. This learning performance is much superior to those exhibited by
humans and more realistic learning algorithms in cross-situational experiments. We show that introduction of discrimination limitations using Weber's law and  forgetting
reduce the performance of the associative algorithm to the human level.
\end{abstract}


\maketitle

\section{Introduction} \label{sec:Intro}

Early word-learning or lexicon acquisition by children, in which the child learns a fixed and coherent lexicon from 
language-proficient adults, is still  a polemic problem in developmental psychology \cite{Bloom_00}. 
The classical associationist viewpoint, which can be traced back to empiricist philosophers such as Hume and Locke, contends that the mechanism of word learning is 
sensitivity to covariation -- if two events occur at the same time, they become associated -- being part of humans' domain-general learning capability. 
An alternative viewpoint, dubbed social-pragmatic theory, 
claims that the child makes the connections between words and their referents by understanding the referential intentions of others. This idea, which seems to be originally 
due to Augustine,  implies that children use their intuitive psychology or theory of mind \cite{Adolphs_03}  to ‘read’ the adults' minds. Although a variety of experiments with 
infants demonstrate that they exhibit a remarkable statistical learning capacity \cite{Bates_96},   the findings that the word-object mappings are generated both fast and errorless 
by children are difficult to account for by any form of statistical learning. We refer the reader to the book by Bloom \cite{Bloom_00}  for a review of this 
most controversial and fascinating theme.

Regardless of the mechanisms children  use to learn a lexicon, the issue of how good humans are at acquiring a new lexicon using statistical learning in  controlled experiments  has been tackled recently 
\cite{Yu_06,Yu_07,Kachergis_09,Kachergis_12,Smith_11,Yu_12}. In addition, it has been  conjectured that  statistical learning may be the principal mechanism in the development of pidgin \cite{Fontanari_11}. 
In this  context (pidgin), however, it is necessary to assume that the agents are endowed with some capacity to grasp the intentions of the others as well as to understand nonlinguistic cues,
otherwise one cannot circumvent the  referential uncertainty inherent in a word-object mapping \cite{Quine_60}.

The statistical learning scenario we consider here is  termed cross-situational or observational learning, and it is based on the intuitive idea that one way that a learner can determine the meaning of a word is to find something in common across all observed uses of that word \cite{Pinker_84,Gleitman_90,Siskind_96}. Hence learning takes place through the statistical sampling of the contexts in which a word appears. There are two competing theories about word learning mechanism within the cross-situational scenario, namely, 
 hypothesis testing and associative learning (see \cite{Yu_12} for a review). The former mechanism assumes that the learner builds coherent hypotheses about the meaning of a word which is
then   confirmed or disconfirmed   by evidence \cite{Markman_90,Xu_07,Frank _08,Waxman_09}, whereas  the latter is based essentially on the counting of co-occurrences of word-object statistics
\cite{Sloutsky_07,Yu_08}. Albeit associative learning can be made much more sophisticated than the mere counting of contingencies \cite{Yu_12}, in this contribution 
we focus on the simplistic interpretation of that learning mechanism, which allows the derivation of explicit  mathematical expressions to characterize 
the learner's performance.

Although cross-situational associative learning   has been a very  popular lexicon acquisition scenario since it can be easily implemented and studied
through numerical simulations (see, e.g., \cite{Fontanari_11,Smith_03a,Smith_03b,NN_09}), there were only a few attempts
 to study analytically this learning strategy  \cite{Smith_06,Blythe_10}.  These works considered a minimal model of cross-situational learning, in which 
the one-to-one mapping between $N$ objects and $N$ words must be inferred through the repeated presentation of 
 $C + 1 < N$ objects (the context) together with a  target word, which refers to one of the objects in the context. The co-occurrences between
 objects and words are stored in a confidence matrix, whose integer entries count  how many times an object has co-occurred with a given word
 during the learning process. The meaning of a particular word is then obtained by picking the object corresponding to the greatest confidence value associated 
 to that word, i.e., the object that has co-occurred  more frequently with that word. In this paper, we expand on the work of Smith et al. \cite{Smith_06}  
 and offer analytical expressions for the learning rates of this minimal associative algorithm
for different 
word sampling schemes, see Eqs.\ (\ref{alpha_sw}), (\ref{alpha_r}) and (\ref{alpha_d}).

To assess the relevance of our findings to the efforts on understanding how humans perform on cross-situational learning tasks, we
use Monte Carlo simulations to  compare the performance of the minimal associative algorithm with the performance of humans for short learning times
\cite{Kachergis_09}  and with the performance  of a more elaborated learning algorithm for  long times \cite{Kachergis_12}. 
Our finding that the accuracy of the minimal associative algorithm  is much higher than that observed in the experiments is imputed to the illimited storage and
discrimination capability of the algorithm. In fact, introduction of errors in the discrimination of confidence values according to Weber's law reduces
the performance to a level  below that of  humans. Somewhat surprisingly, introduction of forgetting  acts synergistically  with our prescription for Weber's law
resulting in an increase of   performance that eventually matches the experimental results.

The rest of this paper is organized as follows. In Sect.\ \ref{sec:task} we describe the learning scenario and in Sect.\ \ref{sec:LL} we introduce and
study analytically  the simplest
associative learning scheme for counting co-occurrences of words and objects, in which the words are learned independently.
We consider first the problem of learning a single word and then investigate the effect of using different word sampling schemes for learning
the complete $N$-word lexicon. 
  In Sect.\ \ref{sec:new} we compare the performance of the minimal associative algorithm with the performance exhibited
by adult subjects. To understand the high efficiency of the algorithm we introduce constraints on its storage and discrimination capabilities and show
how the constraint parameters  can be tunned to describe the experimental results.
  Finally, in Sect.\ \ref{sec:disc} we
discuss our findings and present some concluding remarks.

\section{Cross-situational learning  scenario}\label{sec:task}

We assume that there are $N$ objects,  $N$ words and a one-to-one mapping between words and objects.  
To describe the one-to-one word-object mapping, we use the index $i= 1, \ldots, N$ to represent the $N$ distinct objects and the index  $h = 1, \ldots, N$  to represent the $N$ distinct words.
Without loss of generality, we define the correct mapping as that for which the object represented by $i=1$ is named by the word represented by $h=1$, object represented by $i=2$ by  word
represented by $h=2$, and so on. Henceforth  we will refer to the integers $i$ and $h$ as objects and words, respectively,  but we should keep in mind that they are actually labels to those complex entities.

At each learning event, a target  word, say word $h=1$, is selected and then
$C+1$  distinct objects are  selected from the  list of $N$ objects. This set of $C+1$ objects forms a context for the selected word. The correct 
object ($i=1$, in this case) must be present in the context.
The learner's task is to guess which of the $C+1$ objects the word refers to. This is then an ambiguous word learning scenario in which there are multiple object candidates for any word. 

The parameter $C$ is a measure of the ambiguity (and so of the  difficulty) of the learning task. In particular,  in the case $C=N-1$  the word-object mapping is unlearnable.
At first sight one could expect that  learning  would be trivial for $C=0$  since there is no ambiguity, 
but the learning  complexity depends also on the manner the objects are selected to compose
the contexts. Typically, the objects are  chosen randomly and without replacement  from the list of $N$ objects  (see, e.g., \cite{NN_09,Smith_06,Blythe_10}), which for $C=0$ results in a  
learning   error (i.e., the fraction of wrong word-object associations)   
that  decreases exponentially  with learning rate $-\ln \left ( 1 - 1/N \right )$ as  the number of learning trials $t$ increases. This is so because there is a non-vanishing probability that
some words are not selected in the $t$  trials \cite{Blythe_10}.

In order to avoid testing subjects on the meaning of words they  never heard, most experimental studies on word-learning mechanisms use a deterministic word selection procedure
which guarantees that all words  are uttered before the  testing stage, although some words may be spoken more frequently than others \cite{Yu_06,Yu_07,Kachergis_09,Kachergis_12}. Hence
we consider here, in addition to the random selection  procedure, 
a deterministic selection procedure  which guarantees that all $N$ words are selected in $t=N$ trials. For this procedure the case $C=0$ is trivial and the learning error
becomes zero at $t=N$. However,  since  encountering  words whose meaning is  unknown is not a
rare event  in the real world (hence the utility of dictionaries), a non-uniform Zipfian  random selection of words is likely  to be a more realistic sampling scheme for learning  
natural word-referent associations (see, e.g.,  \cite{Blythe_10}).


\section{Minimal Associative Learning Algorithm}\label{sec:LL}

Here we consider one of the earliest mathematical learning models -- the linear learning model \cite{Bush_55}. The basic assumption of this model is that learning
can be modeled as a change in the confidence with which  the learner associates the target word to a certain object in the context. More to the point, this confidence is represented by a matrix
whose non-negative integer entries $p_{hi}$ yield a value for the confidence with which word $h$ is associated to object $i$.
We assume that at the outset
($t=0$) all  confidences are set to zero, i.e.,  $p_{hi} = 0$ with $i,h = 1, \ldots,N$  and whenever  object $i^*$ appear in a context  in companion with target
word $h^*$  the confidence $p_{i^* h^*}$ increases by one unit.
Hence at each learning trial, $C+1$ confidences are updated. Note that this learning algorithm considers reinforcement only.

To determine which  object corresponds to word $h$ the learner simply  chooses the object index $i$ for which $p_{hi}$ is maximum.  In the case of ties,
the learner  selects  one object at random among those that maximize the confidence  $p_{hi}$.
Recalling our definition of the correct word-object mapping in the previous section, the learning algorithm achieves a  perfect performance when  $p_{hh} > p_{hi}$ for all $h$ and 
$i \neq h$.
The learning error $E$ at a given trial $t$ is then given by the fraction of wrong word-object associations.  Note that  we have
$p_{hi} \leq  p_{hh}$ with $ i \neq h$ since object $i=h$ must appear in the contexts of all learning events in which the target word is $h$ (see Sect.\ \ref{sec:task}). 
In this case, the  learning error of  any single word, say $h$, which we denote by  $\epsilon_{sw}$,  is the reciprocal of the number of objects  for which $p_{hi} = p_{hh}$ with $i \neq h$.

Interestingly, it can easily be shown that this very simple and general learning algorithm is identical to the algorithm presented in \cite{Smith_06} which is
based on detecting the intersections of  context realizations in order to single out the set of confounder objects at a given trial $t$. This equivalence
has already been noted in the literature  \cite{Vogt_04} (see also \cite{Smith_11}).
The minimal associative  learning algorithm
can be immediately adapted to incorporate more realistic features, such as finite memory and imprecision in the comparison of magnitudes, whereas the confounder reducing algorithm is
restricted to an ideal  learning scenario.

A salient feature of the minimal associative learning algorithm which allows the analytical study of  its performance is the fact that words are learned independently.  
This  is easily seen by noting that the confidences $p_{hi}, i=1, \ldots, N$ are updated only when the  target word $h$ is selected. 
This means that, aside
from a trivial rescaling of the learning time, our scenario is equivalent to the experimental settings (see Sect.\ \ref{sec:new}) in which $C+1$ target words are presented
together with a context exhibiting $C+1$ objects, with each object associated to one of the target words \cite{Yu_06,Yu_07,Kachergis_09,Kachergis_12}. 
Taking advantage of this feature,  we will  first solve
a simplified version of the cross-situational learning in which a given target  word $h$ (and its associated object $i=h$)  
appears in all learning trials whereas the $C$ other objects (the confounders) that make up the rest of the context vary in each learning trial. Once  the problem of learning a single word  is 
solved (see Sect.\ \ref{sec:single}), we can easily work out the generalization to  learning the whole lexicon (see Sects.\ \ref{sec:random} and \ref{sec:det}).  We will use
$\tau$ to measure the time of the learning trials in the case of single-word learning and $t$ in the  whole lexicon learning case.

\subsection{Learning a single word}\label{sec:single}

Before any learning event has taken place, the target word may be associated to any one of the $N$ objects, so the initial state of the learning error is always equal to 
$\left(N-1\right)/N$. When the first learning event takes place, the target word may be  incorrectly assigned to  the $C$ other confounder objects shown in the context, so the probability 
of error at the  first trial is always equal to $C/ \left( C+1\right)$. 
In the second trial, there are two possibilities:  the probability of error  is unchanged  because
the same context is chosen  or  the probability of error decreases to the value $n/ \left( n+1\right)$ l with $n < C$ because  $n$ confounder objects of the first context appeared again
in the second trial.
The same reasoning allows us to describe the probability of error in any trial given that this probability is known in the previous trial as described next.

As pointed out, the possible error values are $n/ \left( n+1\right)$  with $n=0,1,...,C$. Labeling these values by the index $n$, the probability of error 
at trial $\tau$ can be written as 
\begin{equation}\label{01}
\mathbf{W} \left ( \tau \right ) = \left(w_{C}\left(  \tau \right), w_{C-1}\left( \tau \right), \cdots , w_{1}\left( \tau \right) ,w_{0}\left(  \tau \right) \right)  .
\end{equation}
The time evolution of $\mathbf{W} \left ( \tau \right )$ is  given by the Markov chain
\begin{equation}\label{02}
\mathbf{W} \left ( \tau + 1 \right )  =  \mathbf{W} \left ( \tau \right )  T ,  
\end{equation}
where $T$  is a $\left ( C + 1 \right) \times \left ( C + 1 \right) $ transition matrix whose entries $T_{m n}$ yield the probability that the error at a certain trial is $n/\left ( n + 1 \right )$ 
given that the error  was $m/\left ( m + 1 \right )$ in the previous trial. Clearly,  $T_{m n } = 0$ for   $m < n$ since  the error cannot increase during
the learning stage in the absence of noise.

It is a simple matter to derive $T_{m n }$ for  $m \geq n$  \cite{Smith_06}. In fact, it is given by  the probability that in $C$ choices one selects exactly $n$ of the $m$ confounder objects from
the list of $N-1$ objects. (We recall that the object associated to the  target word is picked with certainty and so the list comprises $N-1$ objects, rather than $N$, and the number of
selections is $C$ rather than $C+1$.) This is given by the hyper-geometric distribution \cite{Feller_68}
\begin{equation}\label{03}
 T_{m n} = \frac{\dbinom{m}{n}\dbinom{N-1-m}{C-n}}{\dbinom{N-1}{C}} 
\end{equation}
for  $m \geq n$ and $T_{m n} = 0$ for  $m < n$. Since the transition matrix is triangular, its eigenvalues $\lambda_{n}$ with $n=0,1,...,C$ are the elements of the main diagonal that correspond to 
transitions that leave the  learning error unchanged, i.e., 
\begin{equation}\label{04}
\lambda_{n} = T_{n n} =  \frac{\dbinom{N-1-n}{C-n}}{\dbinom{N-1}{C}} .
\end{equation}
Note that $\lambda_0 = 1 > \lambda_{n \neq 1}  > 0$ as expected for eigenvalues of a transition matrix.  In addition,  since 
$\lambda_n/\lambda_{n+1} = \left ( N-1 -n \right )/\left ( C -n \right ) > 1$ the eigenvalues are ordered such that
$\lambda_0 > \lambda_1 > \ldots > \lambda_{N-1}$. 

Recalling that the probability vector is known at $\tau = 1$, namely, $\mathbf{W}_{1} = \left ( 1, 0, \ldots, 0 \right )$ we can write
\begin{equation}
\mathbf{W}  \left ( \tau \right ) = \mathbf{W}  \left ( \tau=1 \right )  T^{\tau - 1} .
\end{equation}
Although  it is a simple matter to  write $T^{\tau -1}$ in terms of the right and left eigenvectors of 
$T$,  this procedure does not produce an explicit
analytical expression for $W_n \left ( \tau \right )$ in terms of the two parameters of the model $C$ and $N$, since we are not able to find
analytical expressions for the eigenvectors. However, Smith et al. \cite{Smith_06} have succeeded in deriving a closed  analytical expression for 
$W_n \left ( \tau \right )$ using the  inclusion-exclusion principle of combinatorics \cite{Cameron_94},  
\begin{equation}\label{Wn}
W_n \left ( \tau \right )  = \binom{C}{n} \sum_{i=n}^C \left ( -1 \right )^{i-n}  \binom{C-n}{i-n} \lambda_i^{\tau - 1},
\end{equation}
where $\lambda_i$, given by Eq. (\ref{04}), is the probability that a particular set of $i$ members  of the $C$ confounders in the
first learning episode $\tau = 1$ appear in any subsequent episode. Although the spectral decomposition of $T$ plays no role in the derivation
of Eq.\ (\ref{Wn}) we choose to maintain the notation $\lambda_i$ for the above mentioned probability.

\begin{figure}[!t]
  \begin{center}
\subfigure{\includegraphics[width=0.48\textwidth]{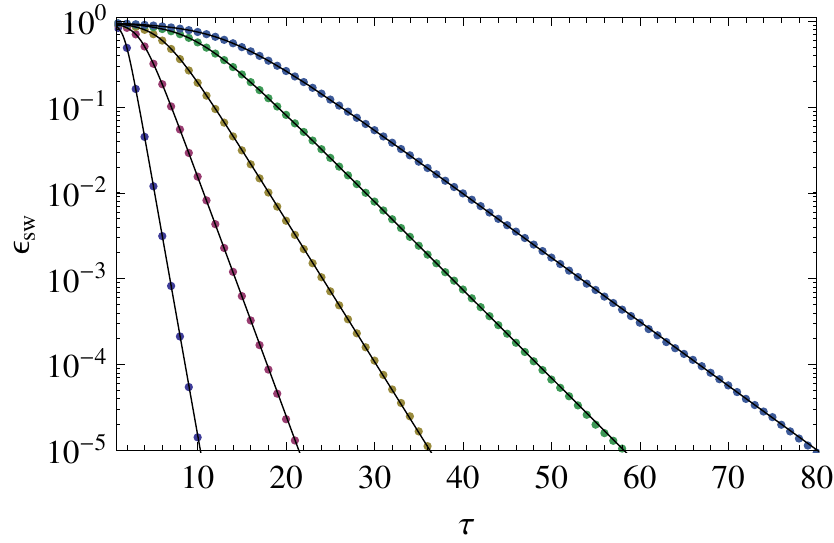}}
 \subfigure{\includegraphics[width=0.48\textwidth]{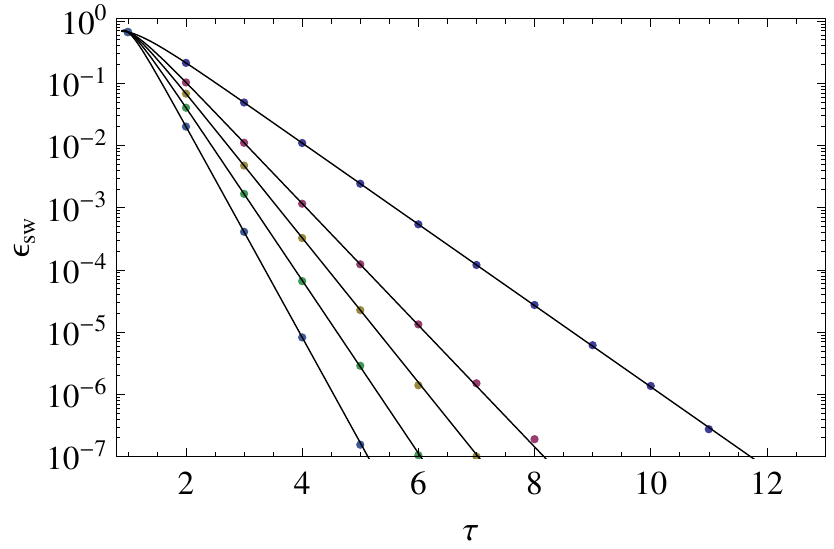}}
  \end{center}
\caption{(Color online)  The expected single-word learning error $\epsilon_{sw}$  as a function of the number of learning trials $\tau$. The solid curves are the results of Eq.\ (\ref{SW1}) and the
filled circles the results of Monte Carlo simulations. The upper panel shows the results for $C=2$ and (left to right) $N$ = 100, 50, 30 and 20, and 	the lower panel for $N=20$  and (left to right) $C$ = 5, 10, 13, 15 and 16.}
\label{fig:1}
\end{figure}

Recalling that a situation described by $n$ corresponds to the learning error   $n/\left ( n + 1 \right )$  we can immediately write the average learning error
for a single word as
\begin{equation} \label{SW1}
\epsilon_{sw}  \left ( \tau \right )  = \sum_{n=0}^C \frac{n}{n+1} W_n \left ( \tau \right ) ,  
\end{equation}
which is valid for $\tau > 0$ only. For $\tau = 0$ one has $\epsilon_{sw}  \left ( 0 \right ) = 1 - 1/N$. 
The dependence of  $\epsilon_{sw}$ on the number of learning trials $\tau$  for different values of  $N$ and $C$ is illustrated in Fig.\ \ref{fig:1} using a semi-logarithmic  scale.  Except for
very small $\tau$, the learning error exhibits a neat exponential decay which is revealed  by considering only the leading non-vanishing contribution to $W_n$ for large
$\tau$, namely,
\begin{equation}\label{SW2}
\epsilon_{sw}  \left ( \tau \right ) \sim \frac{C}{2} \lambda_1^{\tau -1} = \frac{N-1}{2} \exp \left [ -\tau \ln \left ( \frac{N -1}{C} \right ) \right ] .
\end{equation}
Hence the learning rate for single-word learning is 
\begin{equation}\label{alpha_sw}
\alpha_{sw} =  \ln \left [  \left ( N-1 \right ) / C \right ]
\end{equation}
which is zero in the case $C=N-1$, i.e., all objects appear in the context and so
learning is impossible. In the case $C=0$, the learning rate diverges so that $\epsilon_{sw} = 0$ at the first learning trial $\tau = 1$  already.  Most interestingly,  
the learning rate increases with increasing $N$ (see Fig.\ \ref{fig:1}) indicating that the larger the number of objects, the faster the learning of a single word. This apparently 
counterintuitive result has a simple explanation: a  large list of objects to select from  actually decreases  the chances of choosing the same  confounding object  
during the learning events.

\subsection{Learning the whole lexicon with random sampling}\label{sec:random}

We turn now to the original learning problem in which  the learner has to acquire the one-to-one mapping between the $N$ words and the $N$ objects.
 In this section we focus in the case the target word at each learning trial is chosen randomly from the list of $N$ words. Since all words have
the same probability of being chosen, the probability of choosing a particular word is $1/N$.

At trial $t$ we assume that word $1$ appeared $k_1$ times, word $2$ appeared $k_2$ times, and so on  with
$k_1 + k_2 + \ldots +  k_N = t$. The integers $k_i = 0, \ldots, t$ are random variables distributed by the multinomial 
\begin{equation}
P \left(k_{1}, \ldots, k_{N} \right) =  N^{-t} \frac{t!}{k_{1}! \cdots  k_{N}!}   \delta_{t, k_1 + \ldots +  k_N} .
\end{equation}
%

\begin{figure}[!t]
  \begin{center}
\subfigure{\includegraphics[width=0.48\textwidth]{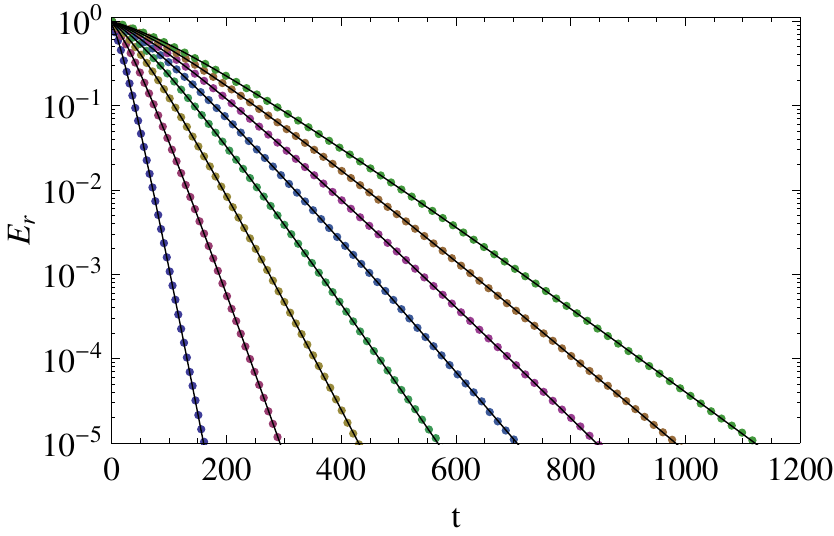}}
\subfigure{\includegraphics[width=0.48\textwidth]{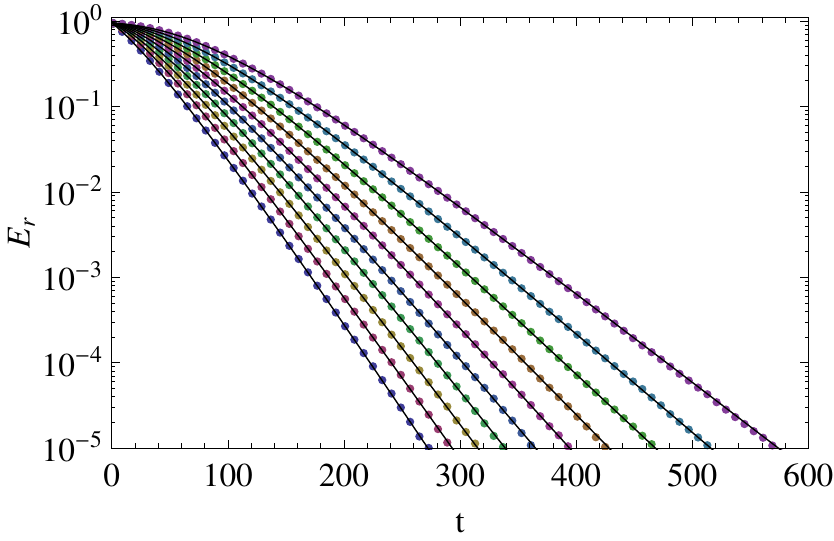}}
  \end{center}
\caption{(Color online)  The expected  learning error $E_r$  in the case the $N$ words are sampled randomly as a function of the number of learning trials $t$. 
The solid curves are the results of Eq.\ (\ref{Er1}) and the
filled circles the results of Monte Carlo simulations. The upper panel shows the results for $C=2$ and (left to right) $N = 10, 20, \ldots, 80$  and 
	the lower panel the results for $N=20$  and (left to right) $C = 1, 2, \ldots, 10 $.}
\label{fig:2}
\end{figure}

Clearly, if word $i$ appeared $k_i$ times in the course of $t$ trials then the expected error associated to it is $\epsilon_{sw} \left ( k_i \right )$ with the
(word independent) single word error given by Eq.\ (\ref{SW1}) for $k_i > 0$. With this observation in mind, we can immediately write the expected learning error in the case the $N$ words are
sampled randomly,
\begin{eqnarray}\label{Er0}
E_{r} \left ( t \right )  &  = &    \sum_{k_1, \ldots, k_N} P \left(k_{1}, \ldots, k_{N} \right)    \frac{1}{N} \sum_{i=1}^N \epsilon_{sw} \left ( k_i \right )  \nonumber \\
& = & \sum_{k=0} ^t \dbinom{t}{k} \left ( \frac{1}{N} \right )^k \left ( 1 - \frac{1}{N} \right )^{t-k} \epsilon_{sw} \left ( k \right ) .
\end{eqnarray}
The sum over $k$ can be easily carried out provided we take into account the fact that $\epsilon_{sw} \left ( k \right ) $ has different prescriptions for the cases $k=0$ and $k > 0$.
We find
\begin{eqnarray}\label{Er1}
E_{r} \left ( t \right )  &  = &  \sum_{n=0}^C \frac{n}{n+1} \dbinom{C}{n} \sum_{i=n}^C  \dbinom{C-n}{i-n} \frac{\left ( -1 \right )^{i-n}}{\lambda_i} \times
\nonumber \\
& &  \left [ \left ( \frac{\lambda_i + N -1}{N} \right )^t   - \left ( \frac{ N -1}{N} \right )^t \right ] \nonumber \\
&   & + \left ( \frac{N-1}{N} \right )^{t+1}
\end{eqnarray}
with $\lambda_i$ given by Eq.\ (\ref{04}). This is a formidable expression which can be evaluated numerically for $C$ not too large and in Fig.\ \ref{fig:2}
we exhibit the dependence of $E_r$ on the number of learning trials for a selection of values of $N$ and $C$.

 To obtain the asymptotic time dependence of $E_r$  we need to keep in the  double sum only the leading order term. Since the summand
 in Eq.\ (\ref{Er1})  vanishes for $n=0$, the largest eigenvalue that appears in that expression is $\lambda_1$, corresponding to the term  $i=n=1$, and so this is the term
 that dominates the sum  in the limit $t \to \infty$.  Hence  $E_r$ exhibits the exponential decay 
 \begin{equation}
 E_r \sim \frac{C}{2 \lambda_1} \left ( \frac{\lambda_1 + N - 1}{N} \right ) ^t = \frac{N-1}{2} \exp \left [ - t \alpha_r \left ( C,N \right ) \right ]
 \end{equation}
where
\begin{equation}\label{alpha_r}
\alpha_r \left ( C,N \right )  = \ln \left [ \frac{ N \left ( N-1 \right )}{C + \left ( N-1 \right )^2} \right ]
\end{equation}
is the learning rate of our algorithm in the case the $N$ words are sampled randomly. As already mentioned, it is interesting that the unambiguous learning scenario $C=0$ 
results in the finite learning rate $ - \ln \left ( 1 - 1/N \right )$ simply because some words may never be chosen in the course of the  $t$ learning trials. 
Interestingly, the learning rate $\alpha_r$ exhibits a non-monotone dependence on $N$ for fixed $C$: for $N > 2C+1$, it   decreases with increasing   $N$  (this is 
the parameter selection used to draw the upper panel of Fig.\ \ref{fig:2}),  and it increases with increasing $N$  otherwise. Recalling that for fixed $C$ the minimum value of $N$ is
$N =C + 1$ at  which $\alpha_r = 0$, increasing  $N$ from this minimal value   must result in an  increase of  $\alpha_r$. The fact that $\alpha_r$ decreases for large $N$ -- an
effect of sampling -- implies that there is an optimal value  $N^* = 2C + 1$ that maximizes the learning speed for fixed $C$. Of course, for fixed $N$ the learning speed is maximized
by $C=0$.

\subsection{Learning the whole lexicon with deterministic sampling}\label{sec:det}
%

\begin{figure}[!t]
\subfigure{\includegraphics[width=0.48\textwidth]{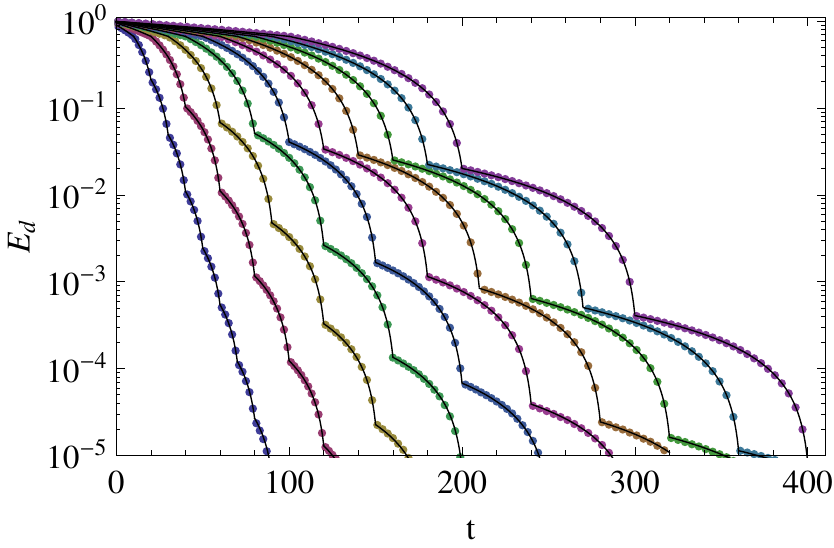}}
 \subfigure{\includegraphics[width=0.48\textwidth]{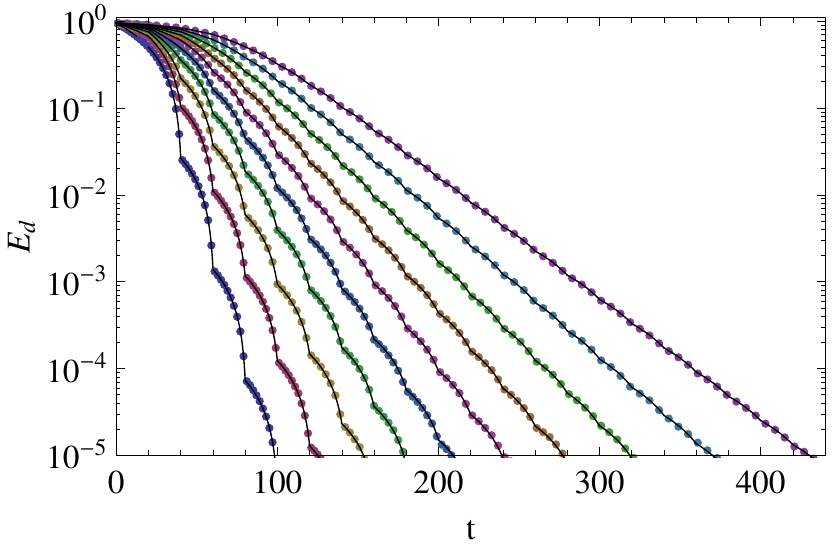}}
\caption{(Color online)  The expected  learning error $E_d$  for the case the $N$ words are sampled deterministically  as a function of the number of learning trials $t$. 
The solid curves are the results of Eq.\ (\ref{Ed1}) and the
filled circles the results of Monte Carlo simulations. The upper panel shows the results for $C=2$ and (left to right) $N = 10, 20, \ldots, 100$  and 
	the lower panel the results  for $N=20$  and (left to right) $C = 1, 2, \ldots, 10 $.}
\label{fig:3}
\end{figure}

To better understand  the effects of the  random sampling of  the $N$ words we consider here a deterministic sampling scheme in which every word is guaranteed 
to be chosen in the course of $N$  learning trials. Let us begin with the first $N$ learning  trials and recall that at time $t=0$ all words have error $\epsilon_{sw} \left ( 0 \right ) =  \left(N-1\right)/N$.
Then during the
learning process  for $t= 1, \ldots, N$  there will be $t$ words with error $\epsilon_{sw} \left ( 1 \right ) = C/ \left(C+1 \right)$ and $N-t$ with error $\epsilon_{sw} \left ( 0 \right ) $ so that the
total learning error for the deterministic sampling is
\begin{equation}\label{Ed0}
 E_d \left( t \right)  = \frac{1}{N} \left[ t  \epsilon_{sw} \left ( 1 \right )  +\left(  N-t \right) \epsilon_{sw} \left ( 0 \right ) \right], \qquad t \leq N . 
\end{equation}
This expression can be easily  extended  for general $t$ by introducing the single-word learning time $\tau = \lfloor t/N \rfloor $, 
\begin{equation}\label{Ed1}
E_d \left( t \right)  = \frac{1}{N} \left[ \left ( t - N \tau \right )  \epsilon_{sw} \left ( \tau + 1 \right )  +
\left(  N \tau + N - t \right) \epsilon_{sw} \left ( \tau  \right ) \right] 
\end{equation}
where $\lfloor x\rfloor $ is the largest integer not greater than $x$. The time-dependence  of the learning error for the deterministic sampling of the
$N$ words is shown in Fig.\ \ref{fig:3}. For $t \gg N$, $\tau$ becomes a continuous variable for any practical purpose, and then  we can see that
$E_d$ decreases exponentially with increasing $t$. Clearly, the learning rate is determined by the single-word learning error [see Eq.\ (\ref{SW2})] and so
replacing $\tau$ by $t/N$ in that equation we obtain
the learning rate for the deterministic sampling case
\begin{equation}\label{alpha_d}
\alpha_d \left ( C,N \right )  = \frac{1}{N} \ln \left [ \frac{  N-1 }{C} \right ] .
\end{equation}
As in the single-word learning case, the learning rate diverges for $C=0$ in accordance with our intuition that in the absence of
ambiguity, the learning task should be completed in $N$ steps. In fact, the learning error decreases linearly with $t$ as given by Eq.\ (\ref{Ed0}).
Similarly to our findings for the random sampling, $\alpha_d$ exhibits a non-monotonic dependence on $N$:  beginning from $\alpha_d = 0$  at
$N=C+1$, it increases  until reaching a maximum at  $N^* \approx e C$ and then decreases towards zero again  as the size of the lexicon  further increases.

It is interesting to compare the learning rates for the two sampling  schemes, Eqs.\ (\ref{alpha_r}) and (\ref{alpha_d}).  In the leading non-vanishing order for large $N$ and $C \ll N$,  we find
$\alpha_r \approx C/N^2$ whereas $\alpha_d \approx \left ( \ln N \right )/N$ . In the more realistic situation in which the context size grows linearly with the lexicon size, i.e.,
$C = \gamma N$ with $\gamma \in \left [0,1 \right ]$,   for large $N$ we find $\alpha_r \approx \left ( 1 - \gamma \right )/N$  and $\alpha_d \approx - \left ( \ln \gamma \right )/N$.
 Hence for small $C$ or $\gamma \approx 0$, the deterministic sampling of words  results in  much faster learning  than the random sampling. For large $C$ or $\gamma \approx 1$, however,
the two sampling schemes produce equivalent results.

\section{Effects of imperfect memory and discriminability}\label{sec:new}

The simplicity of the minimal associative learning algorithm analyzed in the previous section is deceiving. In fact, the algorithm contains two  assumptions that make it
extremely powerful. The first assumption is illimited memory, since
the algorithm  stores the confidence values
from the very first  to the last learning episode, regardless of the number of learning episodes. The second is  perfect discriminability, since 
it always identifies the largest confidence regardless of the closeness to, say, the second-largest one. 

The scheme we use to relax the perfect discriminability assumption is inspired by  Weber's law,  which asserts that the discriminability of two perceived
magnitudes is determined by the ratio of the objective magnitudes. Accordingly,  we assume that the probability that the algorithm selects object $i$ as the
referent  of any given word $h$ is  simply $p_{hi}/\sum_j p_{hj}$, so that  referents  with  similar confidence values  have similar
probabilities of being selected. This differs from the original minimal algorithm for which the referent  selection probability is either
one or zero, except in the case of ties when the probability is divided equally among the referents with identical confidence values.

Forgetting or decaying of the confidence values is implemented  by  subtracting  a fixed factor $\beta \in \left [ 0, 1 \right ]$ from the confidences  $p_{hi}, i=1,\ldots,N $   whenever
word $h$ is absent from a learning episode. The problem with this procedure is that the confidence values may become negative and when this happens  we
reset them to zero. Another difficulty that may rise is when $p_{hi}= 0$ for all $i=1,\ldots,N $ and in this case we reset  $p_{hi}= 1/N$ for all $i=1,\ldots,N $.
These resetting procedures are responsible for  the discontinuities observed in the performance of the algorithm as we will see next. As in the minimal algorithm, 
we  add $1$ to the confidences associated to the target word and the objects exhibited in the context.

Relaxation  of the perfect memory  assumption  makes the forgetting parameter $\beta$ dependent on the  sampling scheme of words, which precludes an 
analytical approach to this problem. As we have  to resort to  simulations  to study
the performance of the modified algorithm anyway, in this section we consider a very specific sampling scheme used in experiments with adult subjects 
to test the effect of varying the frequency  of presentation of the target words on their learning performances \cite{Kachergis_09}. More importantly,
use of this sampling scheme  allows us to 
compare quantitatively the performance of the  minimal as well as of the modified associative learning algorithms with the performances
of the adult subjects.

\begin{figure}[!h]
\centering\includegraphics[width=0.48\textwidth]{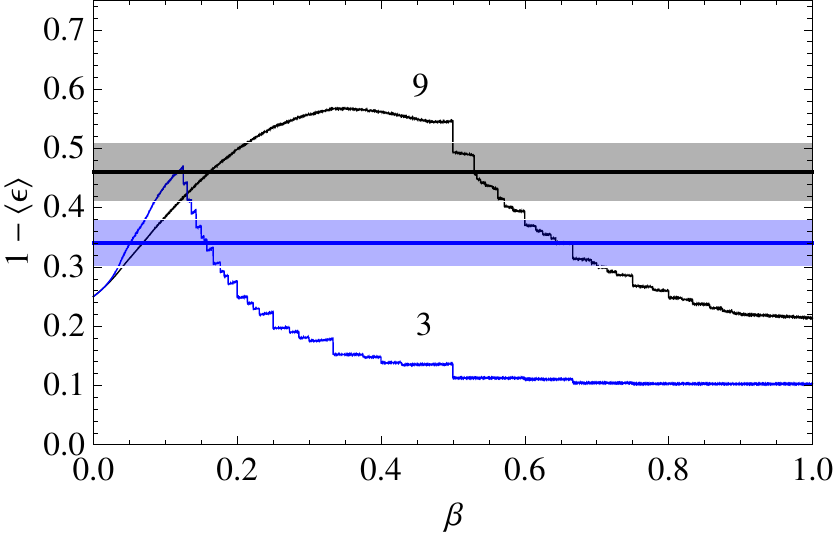}
\caption{(Color online)  Expected accuracy  for the two frequency condition  as function of the forgetting  parameter $\beta$ at learning trial $t=27$.
The curves show the accuracy of the set of words sampled 9 and 3 times as indicated in the figure. The horizontal lines and the shaded zones are
the experimental results \cite{Kachergis_09}. For $\beta \approx 0.16$ we get an excellent agreement between the model and experiments.
}
\label{fig:4}
\end{figure}

\begin{figure}[!h]
\centering\includegraphics[width=0.48\textwidth]{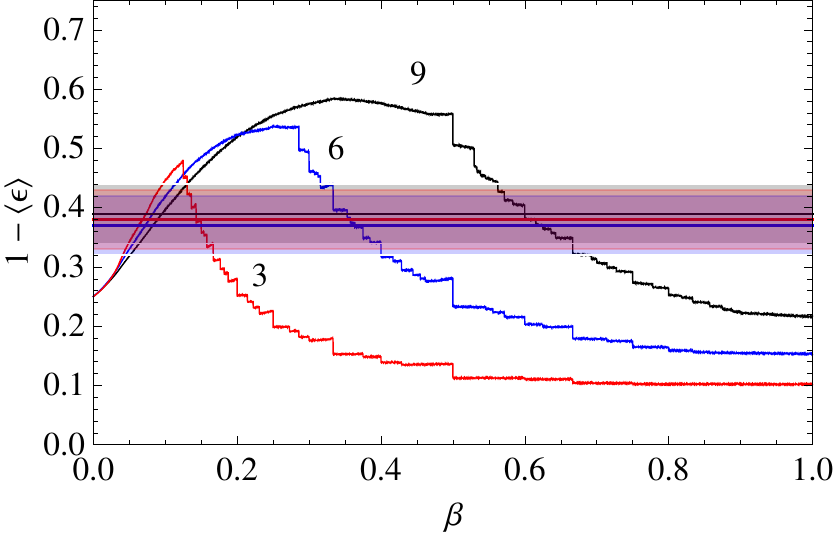}
\caption{(Color online) Expected  accuracy  for the three frequency condition  as  function of the forgetting parameter $\beta$ at learning trial $t=27$.
The curves show the accuracy of the set of words sampled 9, 6  and 3 times as indicated in the figure. The horizontal lines and the shaded zones are
the experimental results  \cite{Kachergis_09}. For $\beta \approx 0.08$ we get an excellent agreement between the model and experiments.
}
\label{fig:5}
\end{figure}

The experiment we consider here aims at evaluating the performance of the associative algorithms in learning a mapping between $N=18$  words and $N=18$  objects  
after 27 training episodes \cite{Kachergis_09}. Each
episode comprises the presentation of $4$ objects together with their corresponding words. Following Ref.\ \cite{Kachergis_09}, we investigate two conditions.
 In the two frequency condition, the $18$ words are  divided into two subsets of  9 words each. In the first subset the 9 words appear 9 times and in the second only 3 times 
 (see Fig. \ref{fig:4}).
 In the three frequency condition,   the $18$ words are divided in three subsets of 6 words each. In the first subset, the 6 words appear 3 times, in the second, 6 times and in
 the third, 9 times  (see Fig. \ref{fig:5}). In these two conditions, the same word was not allowed to appear in two consecutive learning episodes.
 
Once the cross-situational learning scenario  is defined, we carry out $10^4$ runs of  the modified associative learning algorithm  for a fixed
value of the forgetting parameter. The results are shown in terms of the average accuracy $1 - \langle \epsilon \rangle $ as function of $\beta$ in Figs. \ref{fig:4} and \ref{fig:5}.
The horizontal straight lines  and the shaded zones  around them represent the means and standard deviations of the results of experiments carried out with 33 adult subjects  \cite{Kachergis_09}.

Before discussing the interesting dependence of the accuracy on the forgetting parameter exhibited in Figs.\ \ref{fig:4} and \ref{fig:5}, a word is in order about the performance of the original
minimal  algorithm
that is not shown in those figures. In the two frequency condition, the mean accuracy is  $0.99$ for words in the 9-repetition subset and $0.90$ for those in the 3-repetition subset. 
In the three frequency condition, the mean accuracy is $0.99$ for words in the 9- and 6-repetition subsets, and $0.91$ for those in the 3-repetition subset. These accuracy values  are well above those
exhibited in Figs.\ \ref{fig:4} and \ref{fig:5}. Moreover, adding the  forgetting factor to the minimal associative 
algorithm  does not affect its performance, since  subtracting the same quantity from
all confidence values $p_{hi}$ for a fixed word $h$ does not alter the rank order of these confidences.

Although we intuitively expect that  words that appear more frequently would be learned  better, this outcome actually depends on the value
of the forgetting parameter  as shown in Figs.\ \ref{fig:4} and \ref{fig:5}. This counterintuitive finding was first observed in the three frequency condition
experiment on adult subjects \cite{Kachergis_09}. In fact, the results  of those experiments (i.e., the expected accuracies) can be described very well by choosing $\beta = 0.16$
in the two frequency condition and  $\beta = 0.08$ in the three frequency condition.

It is interesting that the choice of a moderate value for  the forgetting  parameter $\beta$  may result in a considerable  improvement of the performance of the algorithm.
This is a direct  consequence of Weber's law prescription for the discrimination  of the confidence values and so there is a  synergy between discrimination and memory
in our algorithm. To see this we note that  at a given learning trial  the ratio between the  probabilities of selecting referent $i=1$ and referent $i=2$  for a  word $h$  
is $r = p_{h1}/p_{h2}$. If word $h$ 
does not appear in the next trial then this ratio becomes 
\begin{equation}
r' = \frac{p_{h1} - \beta}{ p_{h2} - \beta } \approx r \left [ 1  + \frac{\beta}{p_{h1}p_{h2}}  \left ( p_{h1} -p_{h2} \right)  \right ]
\end{equation}
so that $r' > r $ if $p_{h1} > p_{h2}$, thus implying that the forgetting parameter helps the discrimination of the largest confidence. Of course, too large values of
$\beta$ deteriorate the performance of the algorithm as shown in the figures. We note that the dents and jumps in the learning curves are not statistical fluctuations
but consequences of the discontinuities introduced by the ad hoc regularization procedures  discussed before.

The above analysis,  summarized in part by  Figs. \ref{fig:4} and \ref{fig:5},  evinces the better performance of  the  associative algorithm with perfect storage and 
discrimination capabilities when compared with humans' performance for a finite number of learning trials ($t=27$, in the case). In addition,  it shows that 
introduction of imprecision in the discrimination of confidence values following  Weber's law prescription together with forgetting brings that performance down to the  human level. 

For the sake of completeness, it would be interesting to compare the performance of the minimal associative algorithm with humans' performance 
in the limit of  very long learning times, which was in fact the main focus of Sect.\ \ref{sec:LL}. As there are no such experiments  -- we guess it  would be nearly impossible 
to keep the subjects' attention focused on  such  boring tasks for too long -- next we compare the performance of the minimal algorithm with  the performance of 
a rather sophisticated learning algorithm  which, among other things,  models the attention of the learners to 
regular and novel words \cite{Kachergis_12}.   The algorithm is described briefly as follows. 
At any given trial, the confidence values $p_{hi}$ are adjusted according to  the update rule 
\begin{equation}
p_{hi}' = \hat{\beta} p_{hi} + \chi  \frac{  p_{hi} \exp \left [ \lambda \left ( H_h + H_i \right) \right ]}
{\sum_{hi}  p_{hi} \exp \left [ \lambda \left ( H_h + H_i \right) \right ]}
\end{equation}
where  
\begin{equation}
H_h = - \sum_i \Lambda_{hi} \ln \Lambda_{hi}
\end{equation}
with
$\Lambda_{hi} = p_{hi}/\sum_i p_{hi}$, and similarly for $H_i$ with the indexes of the  sums running over the set of words \cite{Kachergis_12}.
In this equation  the entropies $H_h$ and $H_i$  are used as measures of the novelty of word $h$ and object $i$ at the current  learning episode.
The parameter $\hat{\beta}$ governs forgetting, $\chi$ is the weight distributed among the potential associations in the trial, and 
$\lambda$ weights the uncertainty (entropies) and prior knowledge ($p_{hi}$). We refer the reader to Ref.\ \cite{Kachergis_12} for 
a detailed explanation of the algorithm as well as for a comparison with experimental results for short learning times. Here we present its performance in acquiring
the word-object mapping in the simplified  scenario of Sect.\ \ref{sec:LL} (i.e., one target word and $C+1$ objects in the context) for randomly
sampled words.

\begin{figure}[!h]
\centering\includegraphics[width=0.48\textwidth]{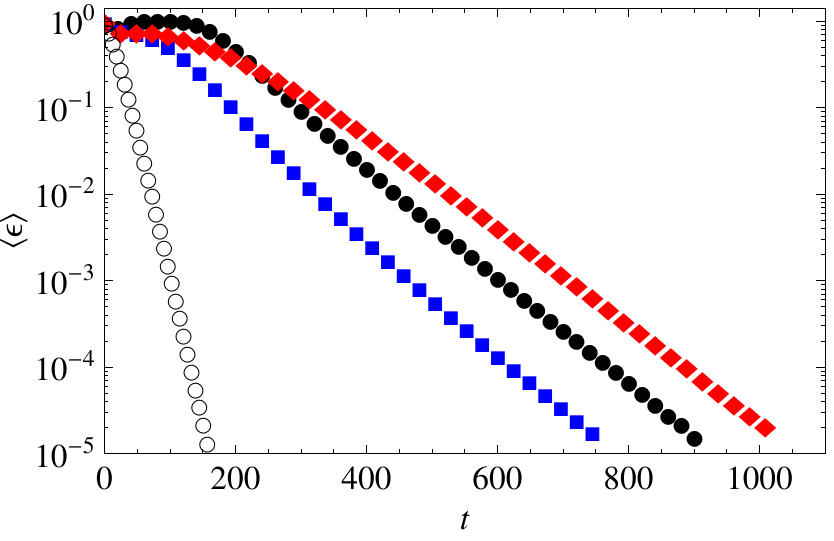}
\caption{(Color online) Expected learning error for $N=10$ and $C=2$ as function of the number of learning trials $t$ in the case words are sampled randomly. 
The open circles are results of the minimal associative algorithm whereas  the filled symbols are the results of the algorithm proposed  by Karchergis et al. \cite{Kachergis_12}:
diamonds ($\chi=3.01, \lambda=1.39, \hat{\beta} =0.64$), circles ($\chi=0.31, \lambda=2.34, \hat{\beta} =0.91$), and  squares ($\chi=0.20, \lambda=0.88, \hat{\beta} =0.96$).
}
\label{fig:6}
\end{figure}

Figure \ref{fig:6} summarizes our findings for $N=10$,  $C=2$ and three selection of the parameter set  ($\chi, \lambda, \hat{\beta}$) used by  Karchergis et al. to 
reproduce the experimental results \cite{Kachergis_12}. The symbols in this figure represent an average over $10^4$ independent samples.
The expected learning error decreases exponentially with increasing $t$ and the rate of learning (the slope of the learning curves
for large $t$ in the semi-log scale) is roughly insensitive to the choice of the parameters of the algorithm. As expected from our previous analysis of short learning times, the minimal associative learning algorithm
performs much better than the more realistic algorithm. These conclusions hold true for a vast variety of different selections of $N$ and $C$, as well as for the deterministic
word sampling  scheme.

\section{Discussion}\label{sec:disc}

As the problem of learning a lexicon within a cross-situational  scenario was studied  rather extensively by Smith et al.  \cite{Smith_06}, 
it is appropriate that we highlight our original contributions to the subject in this concluding section. Although we have borrowed from that work
a key result for the problem of learning a single word, namely,  Eq.\ (\ref{Wn}), even in this case the focal points  of our studies deviate  substantially. In  fact, throughout the
paper our main  goal was the determination of the  learning rates in several learning scenarios, whereas the  main  interest of  Smith et al. was  in quantifying the number of
learning trials required to learn a word with a fixed given probability \cite{Smith_06}. In addition, those authors addressed the problem of the random sampling of words  
 using various approximations, leading to inexact results from where the learning rate $\alpha_r$, see Eq.\ (\ref{alpha_r}), cannot be recovered. As a result,
 the interesting non-monotonic dependence of $\alpha_r$ (and $\alpha_d$, as well) on the size $N$ of the lexicon  passed unnoticed. 
 The study of the deterministic sampling of words  and the introduction and analysis of the effects of limited storage and discrimination capabilities  
 on the original minimal associative algorithm are original contributions of our
 paper.
 
We note that in the cross-situational scenarios studied previously \cite{Smith_06,Blythe_10}  the set of objects that can be associated to a given
word is word-dependent, rather than constant as considered here. In other words, if the target word is $h$ then the elements of the context in a learning episode are 
drawn from a fixed subset of $N_h \leq N $ objects. These subsets can freely overlap with each other. Here we have assumed $N_h = N$ for $h =1, \ldots, N$. 
Of course, this generalization does not affect the analysis of the single-word learning, except that $\epsilon_{sw}$   becomes word-dependent since the parameter
$N$ is replaced by $N_h$ [see Eq.\ (\ref{SW2})]  and similarly for the learning rate $ \alpha_{sw}$ [see Eq.\ (\ref{alpha_sw})]. More importantly, since words are learned independently 
by the minimal  associative algorithm, the single-word learning errors contribute additively to the total lexicon learning error regardless of the sampling procedure
[see Eqs.\ (\ref{Er0}) and (\ref{Ed1})].  Hence
the asymptotic behavior of the total error is determined by the word that takes the longest to be acquired, i.e., the word with the lowest learning rate  or equivalently with the 
smallest  subset cardinality $N_h$.   With this in mind we can easily obtain the learning rates for this more general situation, namely, 
$\alpha_r = \ln \left  \{ N \left ( N_m - 1 \right ) /
\left [ C + \left ( N_m - 1 \right ) \left ( N - 1 \right ) \right ] \right \}$ and $\alpha_d = \ln \left [ \left ( N_m - 1 \right )/C \right ]/N$ where
$N_m = \min_h \left \{N_h, h=1, \ldots, N \right \}$. As expected, in the case $N_m = N$ these expressions reduce
to Eqs.\ (\ref{alpha_r}) and (\ref{alpha_d}).

The cross-situational learning scenario considered here, as well as those used in experimental studies, does not account for the presence of external  noise, such as
the effect of  out-of-context target words. This situation can be modeled by introducing   a probability $\gamma \in \left [ 0, 1 \right ]$ that the correct
object is not part of the context so the  target word can be said to be  out of context. Since we have assumed that learning is based
on the perception of differences in the co-occurrence of objects and target words, in the case all $N$ objects have the
same probability of being selected to form the contexts regardless of the target word, such a purely observational learning is clearly unattainable.
To determine the critical value of the noise parameter $\gamma_c$ at which this situation occurs we simply equate  the
probability of selecting the correct object with the probability of selecting any given confounding object to compose
the context in a learning episode,
\begin{equation}\label{G1c}
1 - \gamma_{c} = \frac{\left( 1 - \gamma_{c}\right) C}{N-1} + \frac{\gamma_{c} \left(C+1\right)}{N-1}, 
\end{equation}
from where we get 
\begin{equation}\label{G2c}
\gamma_{c} = 1 -\frac{C+1}{N}.
\end{equation}
Since in this case  all objects and all words are equivalent, in the sense they have the same probability of co-occurrence, the
average single-word  learning error, as wells as  the total error regardless of the sampling scheme,  is simply  $\epsilon_{sw} = 1 - 1/N$.
We refer the reader to Ref.\ \cite{Tilles} for a  detailed study of the behavior of the minimal associative learning algorithm near the critical noise parameter
 using statistical mechanics techniques.
Here we emphasize that the existence of $\gamma_c$ is not dependent on the algorithm used to learn the word-object mapping. Rather, it is a limitation
of cross-situational learning in general.

 The simplifying feature of our model  that  allowed an analytical approach, as well as extremely efficient Monte Carlo simulations  (in all graphs the error bars were smaller than the symbol sizes),
 is the fact that words are learned independently from each other. In this context, the minimal associative algorithm considered here corresponds to the optimal learning strategy. 
Moreover,  the fact that the minimal associative algorithm exhibits effectively illimited storage and discrimination capabilities makes its learning performance much superior to that
of adult subjects in controlled  experiments \cite{Kachergis_09} and to that of sophisticated algorithms designed to capture the strategies used by  humans in the observational  learning task \cite{Kachergis_12}.
Interestingly, introduction of errors in the discrimination of the confidence values using Weber's law reduced the performance of the minimal algorithm 
to the level reported in the experiments. Perhaps, sophisticated 
learning strategies such as the  mutual exclusivity constraint \cite{Markman_90}, which directs children to map novel
words to unnamed referents,  have evolved to compensate the limitations imposed by Weber's law  to evaluate the frequency of co-occurrence of words and referents.

\section*{Acknowledgments}

This research was supported by The Southern Office of
Aerospace Research and Development (SOARD), Grant No.
FA9550-10-1-0006, and Conselho Nacional de Desenvolvimento
Cient\'{\i}fico e Tecnol\'ogico (CNPq). P.F.C.T. was supported by  
Funda\c{c}\~ao de Amparo \`a Pesquisa do Estado de S\~ao Paulo 
(FAPESP).


\end{document}